\documentclass{sig-alternate}
\newcommand{\commentout}[1]{}
\commentout{
\newtheorem{THEOREM}{Theorem}[section]
\newenvironment{theorem}{\begin{THEOREM} \hspace{-.85em} {\bf :} }%
                        {\end{THEOREM}}
\newtheorem{LEMMA}[THEOREM]{Lemma}
\newenvironment{lemma}{\begin{LEMMA} \hspace{-.85em} {\bf :} }%
                      {\end{LEMMA}}
\newtheorem{COROLLARY}[THEOREM]{Corollary}
\newenvironment{corollary}{\begin{COROLLARY} \hspace{-.85em} {\bf :} }%
                          {\end{COROLLARY}}
\newtheorem{PROPOSITION}[THEOREM]{Proposition}
\newenvironment{proposition}{\begin{PROPOSITION} \hspace{-.85em} {\bf :} }%
                            {\end{PROPOSITION}}
\newtheorem{DEFINITION}[THEOREM]{Definition}
\newenvironment{definition}{\begin{DEFINITION} \hspace{-.85em} {\bf :} \rm}%
                            {\end{DEFINITION}}
\newtheorem{CLAIM}[THEOREM]{Claim}
\newenvironment{claim}{\begin{CLAIM} \hspace{-.85em} {\bf :} \rm}%
                            {\end{CLAIM}}
\newtheorem{EXAMPLE}[THEOREM]{Example}
\newenvironment{example}{\begin{EXAMPLE} \hspace{-.85em} {\bf :} \rm}%
                            {\end{EXAMPLE}}
\newtheorem{REMARK}[THEOREM]{Remark}
\newenvironment{remark}{\begin{REMARK} \hspace{-.85em} {\bf :} \rm}%
                            {\end{REMARK}}
}                      {}

\newtheorem{theorem}{Theorem}[section]
\newtheorem{corollary}{Corollary}[section]
\newtheorem{lemma}{Lemma}[section]
\newtheorem{proposition}{Proposition}[section]
\newtheorem{definition}{Definition}[section]
\newcommand{\thm}{\begin{theorem}}
\newcommand{\lem}{\begin{lemma}}
\newcommand{\pro}{\begin{proposition}}
\newcommand{\dfn}{\begin{definition} \rm}
\newcommand{\rem}{\begin{remark}}
\newcommand{\xam}{\begin{example}}
\newcommand{\cor}{\begin{corollary}}
\newcommand{\prf}{\begin{proof}}
\newcommand{\ethm}{\end{theorem}}
\newcommand{\elem}{\end{lemma}}
\newcommand{\epro}{\end{proposition}}
\newcommand{\edfn}{\bbox\end{definition}}
\newcommand{\erem}{\bbox\end{remark}}
\newcommand{\exam}{\bbox\end{example}}
\newcommand{\ecor}{\end{corollary}}
\newcommand{\eprf}{\end{proof}}
\newcommand{\beqn}{\begin{equation}}
\newcommand{\eeqn}{\end{equation}}
\newcommand{\bbox}{\vrule height7pt width4pt depth1pt}
\newcommand{\clm}{\begin{claim}}
\newcommand{\eclm}{\end{claim}}
\newcommand{\info}{{\mbox{\it info}}}
\renewcommand{\S}{{\cal S}}
\newcommand{\A}{{\cal A}}
\newcommand{\B}{{\cal B}}
\newcommand{\C}{{\cal C}}
\newcommand{\D}{{\cal D}}
\newcommand{\E}{{\cal E}}
\newcommand{\R}{{\cal R}}
\newcommand{\I}{{\cal I}}
\newcommand{\Dom}{{\it DOM}}
\newcommand{\removeTemporarily}[1]{}
\newcommand{\remove}[1]{}
\newcommand{\removeforJoe}[1]{}

\newcommand{\round}{\mathrm{round}}
\newcommand{\union}{\cup}

\begin{document}
\conferenceinfo{STOC'04}{June 13--15, 2004, Chicago, Illinois, USA}
\CopyrightYear{2004}
\crdata{1-58113-852-0/04/0006}

\title{Rational Secret Sharing and Multiparty Computation:\\
Extended Abstract}

\numberofauthors{2}
\author{
\alignauthor Joseph Halpern\titlenote{Work supported in part by NSF
under grant
CTC-0208535, by ONR under grants  N00014-00-1-03-41 and
N00014-01-10-511, by the DoD Multidisciplinary University Research
Initiative (MURI) program administered by the ONR under
grant N00014-01-1-0795, and by AFOSR under grant F49620-02-1-0101.}\\
       \affaddr{Department of Computer Science}\\
       \affaddr{Cornell University}\\
       \affaddr{Ithaca, NY 14853}\\
       \email{halpern@cs.cornell.edu}
\alignauthor Vanessa Teague\titlenote{Supported by OSD/ONR CIP/SW
URI ``Software Quality and Infrastructure Protection for Diffuse
        Computing'' through ONR grant N00014-01-1-0795.}\\
       \affaddr{Department of Computer Science}\\
       \affaddr{Stanford University}\\
       \affaddr{Stanford, CA 94305-9025}\\
       \email{vteague@cs.stanford.edu}
}
\date{Mar 14 2004}
\maketitle
\begin{abstract}
We consider the problems of secret sharing and multiparty
computation, assuming that agents prefer to get the secret
(resp.,~function value) to not getting it, and secondarily, prefer
that as few as possible of the other agents get it.  We show that,
under these assumptions, neither secret sharing nor multiparty
function computation is possible using a mechanism that has a
fixed running time.  However, we show that both are possible using
randomized mechanisms with constant expected running time.
\end{abstract}




\section{Introduction}

{\em Secret sharing\/} is one of the main building blocks in the
modern cryptographic literature.  Shamir's secret-sharing scheme
\cite{shamir} allows someone to share a secret $s$ (a
natural number) among $n$ other agents, so that any $m$ of them
may reconstruct it.  The idea is simple: agent 0, who wants to
share the secret, chooses an $m-1$ degree polynomial $f$ such that
$f(0) = s$, and tells agent $i$ $f(i)$, for $i = 1, \ldots, n$;
$f(i)$ is agent $i$'s ``share'' of the secret.  Any $m$ of agents
$1, \ldots, n$ can recover the secret by reconstructing the
polynomial (using Lagrange interpolation).  However, any subset of
size less than $m$ has no idea what the secret is.

The story underlying this protocol is that, of the $n$ agents, at
most $n-m$ are ``bad''.  While the bad agents might not cooperate,
the good agents will follow the protocol and pool their shares of
the secret. The protocol guarantees that the bad agents cannot
stop the good agents from reconstructing the secret. While for
some applications it makes sense to consider ``good'' agents and
``bad'' agents, for other applications it may make more sense to
view the agents, not as good or bad, but as rational individuals
trying to maximize their own utility.  The agents have certain
preferences over outcomes and can be expected to follow the
protocol if and only if doing so increases their expected utility.

As we show, if we make rather minimal assumptions about the
preferences of the agents, and further assume that the way agents
pool their shares of the secret is by (simultaneously)
broadcasting a message with their share, then there is a problem
with Shamir's secret-sharing scheme: rational agents will simply
not broadcast their shares.
Suppose that each of the agents would prefer getting the secret to
not getting it; a secondary preference is that the fewer of the
other agents that get it, the better. It is then not hard to see
that no agent has any incentive to broadcast his or her share of
the secret. Consider agent 1's situation: either $m-1$ other
agents broadcast their share, or they do not.  If they do, then
agent 1 can reconstruct the secret; if not, she cannot. Whether or
not she sends her share does not affect whether others send theirs
(since all the broadcasts are supposed to happen simultaneously).
Moreover, if only $m-1$ other agents broadcast their shares, then
sending her share will enable others to figure out the secret.  So
if she does not send her share in this circumstance, then she will
be able to figure out the secret (her share combined with the
$m-1$ others will suffice),
while no one who sent their share will. Thus, in game-theoretic
terminology, not sending her share {\em weakly dominates\/}
sending her share.  Intuitively, there is no good reason for her
to send her share. Thus, rational agents running Shamir's protocol
will
not send any messages.

Our first result shows that this problem is not confined to
Shamir's
protocol.  Roughly speaking, we show that, for any mechanism%
\footnote{A mechanism can be thought of as a recommended protocol
for agents to follow, from which they may defect.} for
shared-secret reconstruction with a commonly known upper bound on
the running time, repeatedly deleting
all weakly-dominated strategies
results in a strategy that is equivalent to each agent doing
nothing.  Roughly speaking, we argue that no agents will send a
message  in the last round, since they have no incentive to do so.
Then we proceed by backward induction to show that no agents will
send a message $k$ rounds before the end, for each $k$.
(The actual backward induction process is more subtle, since we
have to argue that, at each step in the deletion process, enough
strategies have not been deleted to show that a strategy we would
like to delete is in fact dominated by another strategy.)
Readers familiar with repeated prisoners'
dilemma will recognize that the argument is similar in spirit
to the argument that shows that rational agents will always defect in
repeated
prisoners' dilemma where the number of repetitions is commonly
known.
In contrast to this impossibility result, we show that there is a
randomized secret-sharing mechanism for rational agents, where the
recommended strategy
is a Nash equilibrium that survives iterated
deletion of weakly-dominated strategies.

We next consider multiparty computation. In the traditional
multiparty computation problem, there is a set of participants,
each of whom has a secret input. The aim of the protocol is to
compute some function of these inputs without revealing any
information other than the function's output, just as if a trusted
party had performed the computations on the agents' behalf. For
example, the secrets could be each person's net worth and the
function would return who is richest.  The protocol should compute
this without revealing any other information about the
participants' wealth.  (This example is known as the {\em
millionaire's problem}, and was first discussed by Yao
\cite{yao:sc}.)  Again, it is
assumed that some of the
parties may be ``bad'', usually less than $1/3$ or $1/2$ of the
total participants, depending on assumptions
\cite{bgw,canetti96studies,ccd,goldreich03,gmw87,yao:sc}. Everyone
else is assumed to be good and to execute the protocol exactly as
instructed.

As in the case of secret sharing, we would like to consider what
happens if the parties are all trying to maximize their utility,
rather than being ``bad'' and ``good''.
A number of new subtleties arise in multiparty computation. As is
well known \cite[Section 7.2.3]{goldreich03}, there is no way to
force parties to participate in a protocol.  We deal with this
problem by assuming that the parties' utilities are such that it
is in their interest to participate
if the protocol is run correctly.  A more serious problem is that
there is no way to force a party to the protocol to use their
``true'' input. A party can correctly run the protocol using an
arbitrary input, which may not necessarily be the same as its true
input. For example, suppose that
each agent has a private bit and the goal is to compute the
exclusive or of the bits.  If agent 1 lies about her bit and
everyone else tells the truth, then agent 1 will be able to
compute the true exclusive or from the information provided by
the trusted party, while no one else will. In some cases, if there
is a trusted party, it may make sense to assume that everyone will
truthfully reveal private information. For example, suppose that
there is a vote, where the candidate with the most votes wins.  In
this case,
almost by definition, what someone
says her vote is is her actual vote.  By way of contrast, consider
a senator trying to determine whether a bill will pass by asking
other senators how they intend to vote.  Then there clearly may be
a difference between how senators say they will vote and how they
actually vote.

Shoham and Tennenholtz \cite{st:bool} characterize which Boolean
functions can be computed by rational agents with a trusted party.
In their model, each agent has a secret input, and everyone is
trying to compute some function of the inputs.
There is a trusted party who waits to be told each player's input,
then computes the value of the function and tells all players.
Every agent's first priority is to learn the true value of the
function; the second priority is to prevent the others from doing
so. Agents may refuse to participate, or they may lie  to the
trusted party about their value.
They call functions for which it is an equilibrium to tell the
truth {\em non-cooperatively computable} (NCC).

Our interest here is in which functions can be computed without a
trusted party.
We show that there is no mechanism with a commonly-known upper
bound on running time for the multiparty computation of any
nonconstant function.  This result is of particular interest since
all the standard multiparty computation protocols do have a
commonly-known upper bound on running time
\cite{bgw,canetti96studies,ccd,goldreich03,gmw87,yao:sc}.
The result also applies to protocols for the fair exchange of
secrets, which is a particularly appropriate case for assuming the
parties are both selfish.  Again all the protocols we could find
have a commonly-known upper bound on the running time \cite{Blum83, BN00,
Cleve89, Damgard95, EGL, LMR83}. As in
the case of secret sharing, we also have a positive result
for multiparty computation. There are multiparty computation
protocols (e.g., \cite{gmw87}) that use secret sharing as a
building block.
By essentially replacing their use of deterministic secret-sharing
by our randomized secret-sharing protocol,
we show that for all NCC functions, we can find
a multiparty computation mechanism
where the recommended
strategy
is a Nash equilibrium that
survives iterated deletion of weakly-dominated strategies.
These results can be viewed as steps in the program advocated in
\cite{FS02} of
unifying the strategic model and computational model in
distributed algorithmic mechanism design.
Our work is related to \cite{McGrewBGW}, but uses a different
solution concept.

The rest of this paper is organized as follows.  In
Section~\ref{sec:Nash} we give the relevant background on Nash
equilibrium, iterated deletion of weakly dominated strategies, and
mechanisms.  In Section~\ref{sec:secretsharing} we consider secret
sharing, sketch the proof of the impossibility result, and give the
randomized  secret-sharing mechanism.  In Section~\ref{sec:multiparty},
we consider multiparty computation.  We conclude in
Section~\ref{sec:conc} with some open problems.

\section{Nash equilibrium, iterated deletion, and
mechanisms}\label{sec:Nash}

We adapt the standard definition of game trees from the game
theory literature slightly for our purposes.
A game $\Gamma$ for $n$ players is described
by a (possibly infinite) forest of nodes.  Intuitively, the root
nodes of the forest describe the possible initial situations in
the game, and the later nodes describe the results of the players'
moves.  We assume that there is a probability distribution over
the root nodes; this can be thought of as a distribution over
possible initial situations. We assume that at each step, a player
receives all the messages that were sent to it by other players at
the previous step, performs some computation, then sends some
messages (possibly none).  Thus, we are implicitly assuming that
the system is synchronous (players know the time
and must decide what messages to send in each round before
receiving any messages sent to them in that round), communication
is guaranteed, and messages take exactly one round to arrive.
These assumptions are critical to the correctness of the
algorithms we present; we believe that rational secret sharing and
multiparty computation are impossible in an asynchronous setting,
or a setting where there is no upper bound on message delivery
time. At each node, each player has a {\em local state\/} that
describes its history, that is, the sequence of computations
performed, messages sent, and messages received, and when each of
these events happened
and encodes its utility function.
Associated with each {\em run\/}
(i.e., path in the forest that starts at a root and is either infinite
or ends in a leaf)
is a tuple
$(u_1, \ldots, u_n)$ of real-valued utilities; intuitively, $u_i$
is player $i$'s utility if that path is played.
Typically utilities are
associated with leaves of game trees.  For finite trees, we can
identify the utility of a run with the utility of its leaf.
Note that we need to consider infinite runs since a randomized mechanism
may not terminate.

Although it is standard in game theory to assume that exactly one
player moves at each node, we implicitly assume that at each step
all the players move.  In game theory, for each player $i$, the
nodes are partitioned into {\em information sets}.  The nodes in
an information set of player $i$ are, intuitively, nodes that
player $i$ cannot tell apart.  Although we do not explicitly use
information sets here, they are easy to define: player $i$'s
information set at a node $v$ consists of all the nodes $v'$ where
she has the same local state.
With this
choice of information sets, it follows that each player $i$ has {\em
perfect recall}, since she remembers all her previous information sets
and her actions.

A {\em strategy\/} or {\em protocol\/} for player $i$ is a
(possibly randomized) function from player $i$'s
local states to actions.
(In the game theory literature, a strategy is a function
from information sets to actions.  Since we are identifying
local states with information sets, our usage of the term strategy
is equivalent to the standard game theory usage.)
A {\em joint strategy\/} $\vec{\sigma} = (\sigma_1, \ldots,
\sigma_n)$ is a tuple of strategies, one for each player.  Note
that a joint strategy determines a distribution over runs, which
in turn determines an expected utility for each player.  Let
$U_i(\vec{\sigma})$ denote player $i$'s expected utility if
$\vec{\sigma}$ is played.

We use the notation $\vec{\sigma}_{-i}$ to denote a tuple
consisting of each player's strategy in $\vec{\sigma}$ other than
player $i$'s.  We then sometimes abuse notation slightly and write
$(\vec{\sigma}_{-i},\sigma_i)$ for $\vec{\sigma}$. A joint
strategy $\vec{\sigma}$ is a {\em Nash equilibrium\/} if no player
has any incentive to do anything different, given what the other
players are doing. More formally, $\vec{\sigma}$ is a Nash
equilibrium if, for all players $i$ and strategies $\sigma_i'$ of
player $i$, $U_i(\vec{\sigma}_{-i},\sigma_i) \ge
U_i(\vec{\sigma}_{-i},\sigma_i')$.

Although Nash equilibrium is a useful concept, there are many Nash
equilibria that, in some sense, are unreasonable.  As a
consequence, many refinements of Nash equilibrium have been
considered in the game theory literature; these are attempts to
identify the ``good'' Nash equilibria of a game (see, e.g.,
\cite{OR94}).  We focus here on one particular refinement of Nash
equilibrium that is determined by iterated deletion of
weakly-dominated strategies.  Intuitively, we do not want a Nash
equilibrium where some player uses a strategy that is weakly
dominated. This intuition is well illustrated with $m$ out of $n$
secret sharing, with $m < n$.  It is a Nash equilibrium for each
player to send its share.  Nevertheless, although a player does
not do better by not sending her share if all other players send
their share (since everyone will still know the secret), a player
does no worse by not sending her share, and there are situations
where she might do better.

\commentout{
Formally, if $\S_i$ is a set of strategies for player $i$, we say
that a strategy $\sigma \in \S_i$ is {\em weakly dominated with
respect to\/} $\S_1 \times \ldots \times \S_n$ if there is some
strategy $\tau \in \S_i$ such that, for some joint strategy
$\vec{\sigma}_{-i} \in  \S_{-i}$, we have
$U_i(\vec{\sigma}_{-i},\sigma) < U_i(\vec{\sigma}_{-i},\tau)$ and,
for all strategies $\vec{\sigma}'_{-i} \in \S_{-i}$, we have
$U_i(\vec{\sigma}'_{-i},\sigma) \le U_i(\vec{\sigma}'_{-i},\tau)$.
Thus, $\sigma$ is weakly dominated with respect to $\S_1 \times
\ldots \times \S_n$ if there is some strategy
$\tau$ that player $i$ should intuitively always prefer to
$\sigma$, since $i$ always does at least as well with $\tau$ as
with $\sigma$, and sometimes does better (given that we restrict
to strategies in $\S_1 \times \ldots \times \S_n$).
}
Formally, if $\S_j$ is a set of strategies for player $j$, $j = 1,
\ldots, n$, we say
that a strategy $\sigma \in \S_i$ is {\em weakly dominated\/} by
$\tau \in \S_i$ with respect to $\S_{-i}$ if, for some strategy
$\vec{\sigma}_{-i} \in  \S_{-i}$, we have
$U_i(\vec{\sigma}_{-i},\sigma) < U_i(\vec{\sigma}_{-i},\tau)$ and,
for all strategies $\vec{\sigma}'_{-i} \in \S_{-i}$, we have
$U_i(\vec{\sigma}'_{-i},\sigma) \le U_i(\vec{\sigma}'_{-i},\tau)$.
Thus,
if $\sigma$ is weakly dominated by $\tau$ with respect to
$\S_{-i}$
then player $i$ should intuitively always prefer $\tau$ to
$\sigma$, since $i$ always does at least as well with $\tau$ as
with $\sigma$, and sometimes does better (given that we restrict
to strategies in $\S_{-i}$). Strategy $\sigma \in \S_i$ is weakly
dominated with respect to $\S_1 \times \cdots \times \S_n$ if
there is some strategy $\tau \in \S_i$ that weakly dominates
$\sigma$ with respect to $\S_{-i}$.

Let $\Dom_i(\S_1 \times \ldots \times \S_n)$ consist of all
strategies for player $i$ that are
weakly
dominated with respect to $\S_1
\times \ldots\times \S_n$.
Given a game $\Gamma$, let $\S_i^0$ consist of all strategies for
player $i$ in $\Gamma$.  Assume that we have defined
$\S_i^k$, for $i = 1, \ldots, n$, where $\S_i^k$ consists of those
strategies for player $i$ that survive $k$ rounds of iterated
deletion.
Let $\S_i^{k+1} = \S_i^k - \Dom_i(\S_1^k \times \ldots \times
\S_n^k)$. Let $\S_i^\infty = \cap_k \S_i^k$.
Thus, $\S_i^\infty$ consists of all those strategies for $i$ that
survive (an arbitrary number of rounds of) iterated deletion of
weakly-dominated strategies.

Note that we are requiring that all weakly-dominated strategies are
deleted at each step.  If we allow an arbitrary subset of
weakly-dominated strategies to be deleted at each step, then which
strategies survive iterated deletion is quite sensitive to exactly which
strategies are deleted at each step.  Deleting all possible strategies
at each step is not only the most natural approach, but the only one
consistent with the intuitions underlying iterated deletion \cite{BK00}.

We take a {\em mechanism\/} to be a pair $(\Gamma, \vec{\sigma})$
consisting of a game and a joint strategy for that game.
Intuitively, a mechanism designer designs the game $\Gamma$ and
recommends that player $i$ follow $\sigma_i$ in that game.  The
expectation is that a ``good'' outcome will arise if all the
players play the recommended strategy in the game.  Designing a
mechanism essentially amounts to designing a protocol; the
recommended strategy is the protocol, and the game is defined by
all possible deviations from the protocol.
$(\Gamma, \vec{\sigma})$ is a {\em practical
mechanism\/} if $\vec{\sigma}$ is a  Nash
equilibrium of the game $\Gamma$
that survives iterated deletion of weakly-dominated strategies.

\commentout{
Our solution concept is very closely related to \emph{dominance
solvability} (\cite{OR94}), which requires that all players are
indifferent between all outcomes that survive the iterated
deletion procedure described above.  Indeed, our impossibility
result for games with a commonly-known upper bound on the running
time shows that such games are dominance solvable and that the
only surviving outcomes are those where no-one sends any
information.
}

\section{Secret Sharing}\label{sec:secretsharing}
In this section, we prove that there is no practical mechanism for
secret sharing with a commonly-known bound on its running time,
provided we make some reasonable assumptions about the preferences
of players, and then show that, under the same assumptions, there
is a randomized practical mechanism for secret sharing that has
constant expected running time.

\subsection{The Impossibility Result}\label{sec:impossibility1}
In this section, we assume for simplicity that there is a share
``issuer'' that can issue secret shares that are atomic and cannot
be subdivided.
The issuer  authenticates the shares, so a player cannot substitute a
false share
for its true one.
We assume that the utility of a run of a mechanism depends only on
which players can compute the secret. Formally, given a run $r$ in
the game tree, let $\info(r)$  be a tuple $(s_0, \ldots, s_n)$,
where $s_i$ is 1 if player $i$ learns the secret in $r$, and is 0
otherwise; let $\info_i(r) = s_i$. The following assumption says
that player $i$'s utility depends just on the information that
each of the players get:
\begin{enumerate}
\item[U1.] $u_i(r) =  u_i(r')$ if
$\info(r) = \info(r')$.
\end{enumerate}
The atomicity assumption implicit in U1 is dropped in
Section~\ref{sec:multipartyimp}, where the utility of a run is
allowed to depend on whatever partial information the players
have.  Note that even in this section we allow the issuer to issue
a sequence of shares.  That is, player $i$ receives shares
$h_{i1}, \ldots, h_{iN}$, and
if $i$ has $m$ of the shares of the form $h_{jN'}$ for some $N'
\in \{1, \ldots, N\}$, then $i$ can compute the secret.  If, for
all $N' \in \{1, \ldots, N\}$, $i$ has fewer than $m$ shares of
this form, then $i$ cannot compute the secret.

The next two requirements encode the assumption that each player
prefers getting the secret to not getting it, and prefers that
fewer of the others get it.
\begin{enumerate}
\item[U2.] If $\info_i(r) = 1$ and $\info_i(r') = 0$, then
$u_i(r) > u_i(r')$.
\item[U3.] If $\info_i(r) = \info_i(r')$, $\info_j(r) \le
\info_j(r')$ for all $j \ne i$, and there is some $j$ such that
$\info_j(r) < \info_j(r')$, then $u_i(r) > u_i(r')$.
\end{enumerate}

Suppose that in run $r$ the players in $P$ learn the secret, while in run
$r'$ the players in $P'$ learn the secret, where either $i \in P \cap
P'$ or $i \notin P \cup P'$.  While it follows from U3 that $u_i(r)
\ge u_i(r')$ if $P \subseteq P'$, we make no assumptions about the
relative utility to $i$ of $r$ and $r'$ if $|P| \ge |P'|$.  It could be,
for example, that there is some particular player $j$ such that $i$
particularly does not want $j$ to learn the secret.

\thm\label{thm:impsec} If players' utilities satisfy U1--U3, then
there is no practical mechanism $(\Gamma,\vec{\sigma}^*)$ for $m$ out of
$n$ secret sharing such that $\Gamma$ is finite and, using
$\vec{\sigma}^*$, some player
learns the secret.  \ethm

The basic idea of the proof is just the backward induction
suggested in the introduction.
Given a mechanism $M =
(\Gamma,\vec{\sigma}^*)$ for $m$ out of $n$ secret sharing, say
that a strategy $\sigma_i$ for player $i$ {\em reveals useful
information at a node $v$\/} in the game tree for $\Gamma$ if (a) $v$
is reachable with $\sigma_i$ (that is, there is some strategy
$\vec{\sigma}_{-i}$ for the other players such
that $(\vec{\sigma}_{-i},\sigma_i)$ reaches $v$ with positive
probability), and (b) according to strategy $\sigma_i$,
at $v$, with positive probability player $i$ sends
some other player $j$
such that $i$ does not know $j$ already has $m$ shares
a share of the secret that $i$ does not know that $j$ already has.
Note that here and elsewhere,
when describing the strategies, we often use phrases such as
``player $i$ knows $P$'' for some proposition $P$.
Player $i$ knows
$P$ at a node $v$ in a game tree if at all nodes $v'$ in the
game tree where $i$ has the same local state,
$P$ is true.  (This usage is
consistent with the standard usage of knowledge in distributed
systems \cite{FHMV}.)  Typically, these statements about knowledge
reduce to concrete statements about messages being sent and
received.  For example, $i \neq j$ knows that $j$ has $k$'s share
of the secret if and only if
\begin{itemize}
  \item $j=k$, or
  \item $i$ has sent $k$'s share to $j$, or
  \item $i$ has received $k$'s share from $j$, or
  \item $i$ has received  a message containing $k$'s share signed
  with $j$'s unforgeable signature.
\end{itemize}
With regard to the last point, note that we allow the possibility of
protocols that make use of unforgeable signatures.  (It would actually
considerably improve the argument if we did not allow them.)
Given a node $v$ in the game tree of $\Gamma$, define $\round(v) = h$ if
there is a path
of length $h$ from $v$ to a leaf
in the game tree and there are no
paths of length $h+1$ from $v$ to a leaf in the game tree.  Thus,
$\round(v)
= 0$ if $v$ is a leaf, and
$\round(v) = \infty$ if there is an infinite path starting at $v$.

Let $\B_i^h$ consist of all
strategies for player $i$ in game $\Gamma$ that reveal useful
information at a node $v$ such that $\round(v) = h$.  (Note that
if $\Gamma$ has no finite paths, then $\B_i^h = \emptyset$.)
Recall from Section~\ref{sec:Nash} that $\S_i^h$
consists of the strategies for player $i$ that survive $h$ rounds
of iterated deletion
(so that $\S_i^0$ consists of all strategies for player $i$). Let
$\R_i^0 = \S_i^0$, and let $\R_i^h = \R_i^0 -
\cup_{i=0}^{h-1}(\B_i^h)$. The backward induction argument would
suggest that,
if the game tree for $\Gamma$ is finite, then
$\S_i^h = \R_i^h$. That is, the strategies that
survive $h$ rounds of iterated deletion
in finite games
are precisely those in
which no useful information is revealed in the last $h$ rounds.
While this is essentially true, it is not as obvious as it might
first appear.

For one thing, while it is easy to see that all strategies in
$\B_i^1$ are weakly dominated---$i$ cannot be better off by
revealing information in the last round---these are not the only
weakly-dominated strategies.  Characterizing the weakly-dominated
strategies is nontrivial.  Even ignoring this problem, consider
how the argument that all strategies in $\B_i^h$ are weakly
dominated with respect to the strategies that remain after $h$
rounds of iterated deletion might go.   Let $\sigma_i \in \B_i^h$,
so that, with positive probability, $i$ reaches a
node $v$ in $\Gamma$
that is no more than $h$ rounds from the end of the game
where $i$ reveals useful information.  Since
$\sigma_i$ has not been deleted earlier, we would expect that $i$
does not reveal useful information at later rounds, nor do any of
the undeleted strategies for the other players.  We would further
expect that $\sigma_i$ would be weakly dominated by the strategy
$\sigma_i'$ that is identical to $\sigma_i$ except that at the
node $v$ and all nodes below $v$, according to $\sigma_i'$, $i$
sends no message (and thus reveals no useful information).  It is
easy to see that $i$ is no worse off using $\sigma_i'$ than
$\sigma_i$---there is no advantage to $i$ in revealing useful
information when no other player will reveal
useful information as a result of getting $i$'s information.
However, to show that $\sigma_i'$ weakly dominates $\sigma_i$, we
must show
that $\sigma_i'$ is not itself deleted earlier, and that there is
some strategy $\vec{\sigma}_{-i}$ for the other players
that was not deleted earlier such that
$U_i(\sigma_i,\vec{\sigma}_{-i}) <
U_i(\sigma_i',\vec{\sigma}_{-i})$.  Intuitively,
$\vec{\sigma}_{-i}$ should be such that $m-2$ other players send
their shares at the same time as $i$, so that with $i$'s share, everyone
can figure out the secret but without it, they cannot.
While this intuition is indeed correct, showing that all the
relevant strategies survive $h$ rounds of iterated deletion turns
out to be surprisingly difficult.
In fact, it seems that we need an almost complete characterization of
which strategies are deleted and when they are deleted in order to prove
the result.  We now provide that characterization.

Besides ``player $i$ knows $P$'', the strategy
descriptions involve phrases such as ``player $i$ considers
$P$ possible'', ``player $i$ tells $j$ that he knows $m$
shares'', ``player $i$ can prove
to $j$ that he (player $i$) knows $m$ shares'', and ``player $i$ can prove to
$j$ that $k$ knows
$m$ shares''.
Possibility is the dual of knowledge,
so that player $i$ considers $P$ possible if player $i$ does not know
$\neg P$.
Player $i$ tells $j$ $P$ or proves $P$ to $j$ at node $v$ if $i$ sends
$j$ messages that guarantee that, at the node after $v$, $j$ knows $P$.
For example,
player $i$ can prove to $j$ that $k$ knows $m$ shares if $i$ can
send to $j$ messages signed with $k$'s unforgeable signature
containing $m-1$ shares other than $k$'s share.

Consider the following families of strategies for player $i$.
Intuitively, these are families of strategies that are deleted in the
iterated deletion procedure.
To simplify the description of these strategies, we write
``[\ldots]'' as
an abbreviation of ``there is a strategy for the other players
such that a node $v$ is reached with positive probability and, at
$v$,''.
We also assume for ease of exposition that secrets are shared just
once.  This is relevant because in the mechanism we present in
Section~\ref{sec:randlength}, secrets may be shared multiple times.
If we consider a sequence of secret sharings, then rather than saying
something like ``player $i$ does not know that player $j$ has $m-1$
shares'', we would have to say ``player $i$ does not know that player
$j$ has $m-1$ shares of a particular sharing of the secret''.

\begin{itemize}
\item
Let $\A_i^1$ consist of all strategies for player $i$ such that
[\ldots]
$i$ has $m$ shares,
$i$ does not know that all the other players have all $m$
shares, and,
with positive probability, $i$ sends each of the other players
enough shares so that, after sending, $i$ will know that they all have
$m$ shares.
(Intuitively, if $i$ already knows the secret, there is no advantage in
$i$ making sure that everyone knows the secret.)

\item If $m=n$, let $(\A_i^1)'$ consist of all strategies for player $i$ such
that [\ldots] $i$ has all $m$ shares and, with positive
probability, $i$ sends out
its share to some player, although $i$ has never previously sent
out its share to any player;
if $m \ne n$, then $(\A_i^1)' = \emptyset$.
(Intuitively, if $i$ knows the secret, and it has information---namely,
its own share---which it has not revealed that is critical to everyone
else learning the secret, then $i$ should not send out this information.
Note that $i$'s share is not critical to others learning the secret if
$m \ne n$, so this condition is vacuous if $m \ne n$.)

\item
If $m=n=2$, let $\A_i^2$ consist of all strategies for player $i$
such that [...] $i$ sends its share to the other player.

If $m =n =3$, let $\A_i^2$ consist of all strategies for player
$i$ such
that [\ldots] $i$ has all three shares,
$i$ considers it possible that some other player $j$ has only its
own share,
$i$ knows that the third player $k$ has all three shares, and,
with positive probability,
either
\begin{itemize}
\item[(a)] $i$ sends $j$ either $i$'s share or $k$'s share, or
\item[(b)] $i$ does not know that $k$ knows that $i$
has all three shares and does not tell $k$ that it has all three
shares.
\end{itemize}
(Intuitively, if $i$ knows that the only player missing the secret
is $j$, then it should try to do what it can to stop $j$ from
getting the secret.  This includes not sending $j$ information and
making sure that that the third player $k$ knows the situation, so
that $k$ will not send $j$ information.)

If $m =3$ and $n =4$,
let $\A_i^2$ consist of all strategies for player $i$ such that
[\ldots] there exist players $j,k,l$ such that the utility to $i$
if $i$, $j$, and $k$ learn the secret is no higher than the
utility to $i$ if $i$, $j$, and $l$ learn the secret, $i$ knows
that everyone knows that everyone has $k$'s share and $l$'s share,
$i$ considers it
possible that $k$ and $l$ lack both $i$'s share and
$j$'s share, and, with positive probability, $i$ sends its share
or $j$'s share to player $k$.
(Intuitively, if $i$ has the secret and knows that $j$ knows the secret,
but considers it possible that both $k$ and $l$ are
missing shares, it should not guarantee that $k$ learns the secret if $k$
learning  the secret is at least as bad as $l$ learning the secret.
While this intuition seems very reasonable,
note that it applies only if $i$ knows that $k$ and $l$ are
missing at most one share.)

If $m =n =4$,
let $\A_i^2$ consist of all strategies for player $i$ such that
[\ldots] there exist players $j,k,l$ such that the utility to $i$
if $i$, $j$, and $k$ learn the secret is no higher than the
utility to $i$ if $i$, $j$, and $l$ learn the secret,
$i$ has all four shares, $i$ knows
that  $j$ has all four shares, and that
everyone knows that everyone has all the shares other than possibly $i$'s,
$i$ considers it possible that $k$ and $l$ both lack
$i$'s share,
and, with positive probability,
$i$ sends its share to
player $k$.
(The intuition here is the same as in the $m=3$, $n=4$ case, but again,
note that it applies only in quite restricted circumstances.)

Otherwise, $\A_i^2 = \emptyset$.

\item Let $(\A_i^2)'$ consist of all strategies for player $i$
such that [\ldots] $i$ has $m$ shares, $i$ considers it possible
that $j$ does not have all $m$ shares, $i$ knows that $j$ has
$m-1$ shares and $i$ can prove this to all the other players, $i$
knows that all players $k \ne j$ have $m$ shares and can prove
this to each player $k' \notin \{i,j,k\}$, and, with positive
probability, $i$ does not prove to each player $k \ne j$ that each
player $k' \notin \{j,k\}$ has $m$ shares
and that $j$ has $m-1$ shares.
(The intuition here is much as that for $\A_i^2$ in the case
$m=n=3$:~if $i$ knows that all but one player has the secret, it
should do what it can to prevent that player from getting the
secret, which includes making sure that the other players know the
situation.)

\item If $m = n=3$, let $\A_i^3$ consist of all strategies for player
$i$ such that [\ldots] $i$ has all three shares,
$i$ considers it possible
that some player
$j$ does not have all three shares, and, with positive
probability, reaches a node $v'$ at the next step such that $i$
knows at $v'$ that
the third player $k$ will eventually know all three shares,
and either
\begin{itemize}
\item[(a)]\label{A2a}
in getting from $v$ to $v'$, $i$ sends $j$ a share that $i$ does not
know at $v$ that $j$ has;
\item[(b)]\label{A2b} $i$ does not know at $v'$ that $k$ has all three
shares;
or
\item[(c)]\label{A2c} $i$
does not know at $v'$ that $k$ knows that $i$ has all three shares.
\end{itemize}
(The situation here is similar to that in the $m=n=3$ case of
$\A_i^2$, except that now, rather than $i$ knowing at $v$ that $k$
has all three shares, $i$ knows only that, with positive probability,
$k$ will have all three shares.  If $i$ knows that
$k$ will eventually have all three shares, then $i$ might as well
tell $k$ all three shares right away, and also tell $k$ that $i$
has all three shares.  This will prevent $k$ from sending $j$
information.)

If $m =3$ and $n =4$,
let $\A_i^3$ consist of two sets of strategies:

(1) All strategies for player $i$ such that
[\ldots] there exist players $j,k,l$ such that
$i$ knows that $j$ is indifferent as to whether $i$, $j$, and $k$ or $i$
$j$, and $l$
learn the secret,
 $i$ knows that $j$ has $k$'s share and $l$'s
share, $i$ knows that everyone (except possibly $j$) knows that
everyone has $k$ and $l$'s shares and can prove
this to $j$, $i$ considers it possible that $k$ and $l$ lack both
$i$'s share and $j$'s share, and, with positive probability, $i$
either
\begin{itemize}
\item[(a)] sends  its share or $j$'s share to $k$
or $l$, or
\item[(b)] does not prove to $j$ that everyone
knows that everyone has
$k$ and $l$'s shares.
\end{itemize}
(The situation here is the same as that in the $m=3$, $n=4$ case
of $\A_i^2$, except that $i$ does not necessarily know that $j$
knows that everyone has $k$ and $l$'s
shares.  By sending the messages, $i$ can ensure that the
antecedent of $\A_j^2$ holds, so that $j$ will not send messages
to $k$.)

(2) All strategies for
player $i$ such that [\ldots] there exist players $j,k,l$ such
that $i$ knows that the utility to $j$ if $i$, $j$, and $k$ learn the
secret is no greater than the utility to $j$ if $i$, $j$, and $l$
learn the secret, $i$ knows that $j$ knows that everyone knows
that everyone has $k$ and $l$'s shares, $i$ considers it possible
that $k$ and $l$ lack both $i$'s share and $j$'s share, and, with
positive probability, $i$ sends  its share or $j$'s share to $l$.
(Here, $i$ knows that $j$ will not send useful information to $k$,
because that would be in $\A_j^2$. Hence sending a third share to
$l$ produces the worst possible outcome for $i$.)

If $m =n =4$,
let $\A_i^3$ consist of all strategies for player $i$ such that
[\ldots] there exist players $j,k,l$ such that the utility to $i$
if $i$, $j$, and $k$ learn the secret is no higher than the
utility to $i$ if $i$, $j$, and $l$ learn the secret,
$i$ has all four shares, $i$ considers it possible that $k$ and $l$ both lack
$j$'s share, with positive probability, $i$ reaches a node $v'$ at the
next step such that, at $v'$, $i$ knows
that eventually $j$
will both know all
four shares and that everyone knows that everyone has all shares except
possibly $j$'s,
but either
\begin{itemize}
\item[(a)] at $v'$, $i$ does not ensure that $j$ knows these facts, or
 \item[(b)] in getting from $v$ to $v'$, $i$ sends $j$'s share to $k$.
\end{itemize}
(The situation here is much like the antecedent of the $m=n=4$
case of $\A_i^2$, except that $i$ does not know at $v$  that $j$
has all four shares or that $j$ knows that everyone has the two
shares other than $j$'s; however, $i$ does know
that, with  positive probability, this will eventually be the case.
Intuitively, if it eventually is going to be the case, then $i$
should make it happen as quickly as possible, since then the
antecedent to $\A_j^2$ will hold, and $j$ will not send messages
to $k$.)

If $m = 2$ or $n > 4$, then $\A_i^3 = \emptyset$.

\item
Let
 $(\A_i^3)''$ consist of all strategies for player $i$ such that
[\ldots] $i$ has $m$ shares, $i$ considers it possible that some
player $j$ does not have all $m$ shares, $i$ knows that $j$
has $m-1$ shares and $i$ can prove this to all the other players,
$i$
knows that all players $k \ne j$ have $m$ shares and there is a
player $k' \notin \{i,j\}$ such that for all $k \notin
\{i,j,k'\}$,
$i$ can prove
 to each player $k'' \notin \{i,j,k\}$
 that $k$ has $m$ shares,
$i$ considers it possible that there is at least one player who
does not know that all players except $j$ have $m$ shares and that
$j$ has $m-1$ shares,
and, with positive probability, $i$ does not provide $k'$ with
evidence that it can use to prove to everyone else that
it ($k'$) knows that
each
player $k \notin \{j,k'\}$ has $m$ shares and that $j$ has $m-1$
shares
or there exists a player $l' \notin \{i,j,k'\}$ such that $i$ can
prove to $l'$ that all players have $m$ shares, except possibly
$j$ who has $m-1$, $i$ considers it possible that $l'$ does not
already know this fact, and $i$ does not prove it to $l'$.
(The situation here is that $i$ knows that everyone has the secret but
$j$, and $j$ only needs one share to get the secret.  In that case, $i$
should make everyone else aware of the situation, as quickly
as possible.)

\item If $m = n =3$, let $(\A_i^3)'''$ consist of all strategies for player $i$
such that [\ldots] $i$ has $j$'s share but not $k$'s,
$j$ has $i$'s share, $i$ does not know both that $k$ has all three
shares and that $k$ knows that $j$ has $i$'s share, and $i$ sends
messages to $k$ that ensure that $k$ has all three shares and
that $k$
knows that $j$ has $i$'s share.
(Intuitively, $i$ should not send information to $k$ that might prevent
$k$ from later sending useful information to $i$.)

If $n > 3$ or $m \ne n$, then $(\A_i^3)''' = \emptyset$.

\item Let $\C^h_i$, consist of all strategies for
$i$ such that [...]
(a) $\round(v) = h$, (b) $i$ knows $m$ shares, (c) if $m=n$,
$i$ has sent out its share earlier, and (d)
with positive probability, $i$ reaches a node $v'$ at the next step such
that there exists a player $j$ such that $i$ knows at $v'$ that all the
players but $j$ know $m$ shares,
in going from $v$ to $v'$, $i$
sends $j$ a share that $i$ does not know at $v$ that $j$ has, and $i$
sends no useful information at any node $v''$ with
$\round(v'') > h$.

\item Let $\D^h_i$ consist of all strategies for $i$
such that [...]
(a) $\mbox{round}(v) = h$,
(b) $i$ knows all the shares,
(c) $i$ has sent out its share earlier, (c) $i$ reveals no useful
information at a node $v'$ with  $\round(v') > h+1$, and (d) there exist
players $j$
and $k$ such that, at $v$, $i$ knows that all the other players
but $j$ and $k$ have $m$ shares and can prove this to $j$ but does
not, $i$ knows that at the step immediately after $v$, $j$ will
have all $m$ shares, and $i$ considers it possible that, after $v$,
$k$ will not have all $m$ shares
  and that $j$ will not know that everyone other than $j$ and $k$
  has $m$ shares.

\item If $m=3$, let $(\D^h)'_i$ consist of all
strategies for player $i$
satisfying (a), (b)
and (c) from the definition of $\D^h_i$, and also
(d$'$) there exist $j$ and $k$ such that, at $v$, all players know $j$ and
$k$'s shares, and $i$ can prove to $j$ that all players besides
$j$ and $k$ know $j$ and $k$'s shares, and either
 $i$ considers it possible that $j$ does not know that all players
 but $j$ and $k$ have $m$ shares, and does not prove this to $j$,
or $i$ considers it possible that $j$ has only two shares, and
sends $j$ a third share.

If $m=n=4$, let $(\D^h)'_i$, consist of all
strategies for player $i$
satisfying (a), (b) and (c) from the definition of $\D^h_i$, and also
(d$''$) there exist players
$j$, $k$, and $l$ such that $i$ can prove to $j$ that $i$ has all
four shares, $i$ knows that $l$ has all four shares and can prove
to $j$ that $l$ has $j$ and $k$'s shares, $i$ knows that $j$ and
$k$ have each other's shares and that $k$ has $l$'s share, and
either (i) $i$ knows that $j$ knows $l$'s share or $i$ sends $l$'s
share to $j$ but does not prove to $j$ that $i$ knows all four
shares and that $l$ knows $j$ and $k$'s shares or (ii) $i$ sends
$j$ $i$'s share if it does not already know that $j$ has $i$'s
share.

Otherwise, $(\D^h)'_i = \emptyset$.
\end{itemize}

Recall that $\B_i^h$
consists of all strategies
for player $i$ that reveal
useful information at a node $v$ with $\round(v) = h$. Let $\A^1 =
\cup_{i=1}^n \A_i^1$; $(\A^1)'$, $\A^2$, $(\A^2)'$, $\A^3$,
$(\A^3)'$, $(\A^3)''$, $(\A^3)'''$,
$\B^h$, $\C^h$, $\D^h$, and $(\D^h)'$, $h = 1, 2, \ldots$ are defined
similarly.
Let $\E^j = \A^j \cup (\A^j)' \cup \B^j \cup \C^{j+1} \cup
\D^{j+1} \cup (\D^{j+1})'$ for $j = 1, 2$; let $\E^3 = \A^3 \cup
(\A^3)' \cup (\A^3)'' \cup (\A^3)''' \cup \B^3 \cup \C^4 \cup \D^4
\cup (\D^4)'$; let $\E^4 = \A^4 \cup \B^4 \cup \C^5 \cup \D^5 \cup
(\D^5)'$; let $\E^j = \B^j \cup \C^{j+1} \cup \D^{j+1}$ for $j
\geq 5$.

\pro\label{pro:delete}
Let $M$ be a mechanism for secret sharing.
After $k$ steps of iterated deletion, all the
strategies in $\E^k$ have been deleted; moreover, no deterministic
strategy not in $\E^1 \cup \E^2 \cup \ldots \cup \E^k$  has been
deleted.
\epro

Note that Proposition~\ref{pro:delete} provides a complete
characterization of when deterministic strategies are deleted, but
does not do so for randomized strategies.  Knowing  when
deterministic strategies are deleted turns out to suffice to prove
the result by induction. It immediately follows from
Proposition~\ref{pro:delete} that there is no practical mechanism
for secret sharing with a
finite game tree: no strategy where any
player sends her share
survives more than $N$ steps of iterated deletion, where $N$ is
a bound on the depth of the game tree.

If $m=n=2$,
all strategies where a player sends its share to another player must be
in $\A^2$.  Thus, the following is an immediate corollary to
Proposition~\ref{pro:delete}.

\cor There is no
practical mechanism for 2 out of 2 secret sharing
(even with an infinite game tree).
\ecor

\subsection{A Randomized Practical Mechanism for Secret Sharing}
\label{sec:randlength}

In light of Theorem~\ref{thm:impsec}, the only hope of getting a
practical mechanism for secret sharing lies in using uncertainty
about when the game will end to induce cooperation.
We now present a randomized protocol for 3 out of 3 secret
sharing, and then show how to extend it to $m$ out of $n$ secret
sharing.

Suppose that players can toss coins in a way that everyone is
forced to reveal their coin tosses after a round is over. Consider
the mechanism whose suggested strategy is as follows:~everyone
tosses their coin, and is supposed to send their secret if their
coin lands heads. In the next step, everyone reveals their coin.
If everyone learns the secret, or if someone cheats (fails to send
their share even though their coin was heads), then the game ends.
Otherwise the issuer issues new shares of the secret (that is,
uses a completely different polynomial and sends shares of that
polynomial), and the process repeats.

Consider the incentives of a player that has tossed heads and is
supposed to send its share. If it withholds its share in the last
step it might be lucky, because it might happen that the other two
players are also about to send their shares. Then it will learn the
secret when the others do not, which it considers the best
possible outcome. However, if the others do not both send their
shares but detect that the first player has cheated, they will
stop the protocol and nobody will learn the secret. This is a
worse outcome than the honest one for the cheater.  This mechanism
ensures that when a player is considering withholding its share
when it ought to send it, the probability of getting caught but
not learning the secret is high ($3/4$), while the probability of
learning the secret when no one else does is only $1/4$.  As long
as $\frac{1}{4} u_i (\mbox{only $i$ learns the secret}) +
\frac{3}{4} u_i(\mbox{no one learns the secret}) <
u_i(\mbox{everyone learns the secret})$, then player $i$ will not
be tempted to cheat.  If player $i$'s utilities do not satisfy
this inequality, the probability of heads can be modified
appropriately.

Unfortunately, this mechanism still has a problem: even if
everyone is honest, there is a chance that one of the players
might learn the secret when the others do not.  If exactly two of
the coins land heads, then the player who tossed tails will be
able to reconstruct the secret, but the other two will not.  The
one who already knows the secret will certainly have no incentive
to continue the game at that point!  We solve this problem by
tossing the coins in such a way that
if exactly two players get heads, then no one learns the secret.
We proceed as follows.

Call the players 1, 2, and 3.%
\footnote{Note that the secret issuer, player 0, is taken to be
honest and is not part of the game.} For $i \in \{1,2,3\}$, let
$i^+$ denote $i+1$ except that $3^+$ is 1; similarly $i^-$ is
$i-1$ except that $1^-$ is 3.  Consider the following protocol:
\begin{enumerate}
  \setcounter{enumi}{-1}
  \item The issuer sends each player a signed share of the secret,
        using 3 out of 3 secret sharing.
\item Each player $i$ chooses a bit $c_i$ such that $c_i = 1$ has probability
$\alpha$ and $c_i = 0$ has probability $1-\alpha$, and a bit $c_{(i,+)}$ at
  random (so that 0 and 1 both have probability $1/2$).  Let $c_{(i,-)} =
  c_i \oplus c_{(i,+)}$.    Player $i$ sends
  $c_{(i,+)}$ to $i^+$ and $c_{(i,-)}$ to $i^-$.  Note that this means that
  $i$ should receive $c_{(i^+,-)}$ from $i^+$ and $c_{(i^-,+)}$ from $i^-$.
  \item Each player $i$ sends $c_{(i^+,-)} \oplus c_{i}$ to $i^-$.
   Thus, $i$ should receive $c_{((i^+)^+,-)} \oplus c_{i^+} = c_{(i^-,-)}
  \oplus c_{i^+}$ from $i^+$.
  \item Each player $i$ computes $p = c_{(i^-,+)} \oplus c_{(i^-,-)} \oplus
 c_{i^+} \oplus c_i = c_{i^-} \oplus c_{i^+} \oplus c_i = c_1 \oplus c_2
   \oplus c_3$.
If $p = c_i = 1$ then player $i$ sends its signed share to the
others.
\item If $p=0$ and $i$ received no secret shares, or if $p=1$ and
              $i$ received exactly one share
(possibly from itself; that is, we allow the case that $i$ did not
 receive any shares from other players but sent its own),
 the issuer is asked to restart the protocol; otherwise, $i$
stops the protocol (either because it has all three shares or
because someone must have been cheating).
\end{enumerate}
If, at any stage, player $i$ does not receive a bit from a player
from whom it is supposed to receive a bit, it also stops the
protocol.

Given a set of possible messages that each player can send at each
point, this protocol determines a mechanism:~there is an infinite
game tree where, at each point, players send some
messages that they are able to send; the
recommended joint strategy is the protocol above. Call
this mechanism $M(\alpha)$, where $\alpha$ is the probability of
$c_i = 1$ at step 1 above.

\thm
\label{thm:rand} For all utility functions satisfying U1--U3, if
$n \ge 3$,
there exists an $\alpha^*$ such that $M(\alpha)$ is a practical
mechanism for $m$ out of $n$ secret sharing for all $\alpha <
\alpha^*$.  Moreover, the expected running time of the recommended
strategy in $M(\alpha)$ is $5/\alpha^3$. \ethm

\prf (Sketch:)
First consider the $m=n=3$ case.
Consider what happens if all the players
follow the protocol.
Player $i$ sends its secret iff $c_i \oplus c_{i^-} \oplus c_{i^+}
= 1$ and $c_i = 1$. This can happen only if $c_1 = c_2 = c_3  = 1$
or if $c_i = 1$ and $c_{i^-} = c_{i^+} = 0$. Thus, all the players
send their shares (and learn the secret) iff $c_1 = c_2 = c_3 =
1$, which happens with probability $\alpha^3$.  If $c_i = 1$ and
$c_{i^-} = c_{i^+} = 0$ (which happens with probability $\alpha
(1-\alpha)^2$), player $i$
sends its share of the secret but the other two players do not, so
no one learns the secret.  If $c_1 \oplus c_2 \oplus c_3 = 0$,
then no player sends its share. Thus, either all players learn the
secret, or no player does. Moreover, the protocol clearly has an
expected running time of $5/\alpha^3$ rounds.

Does player $i$ have
an incentive to cheat at step 3, given that all the other players
follow the protocol?  The  most obvious way that
player $i$ can cheat is by not sending its share when it should,
that is, if $c_i = c_1 \oplus c_2 \oplus c_3 = 1$.  Player $i$
gains in this case if $c_{i^+} = c_{i^-} = 1$, which happens with
conditional probability $\alpha^2/(\alpha^2 + (1-\alpha)^2)$, and
loses if $c_{i^+} = c_{i^-} = 0$, which happens with conditional
probability $(1-\alpha)^2/(\alpha^2 + (1-\alpha)^2)$.  Note that
nothing that player $i$ can do can influence these probabilities,
since each player $j$ chooses its bit $c_j$ independently.  Thus,
a rational player $i$ will cheat only if
\begin{equation}\label{eq1}
\begin{array}{ll}
&\frac{\alpha^2}{\alpha^2 + (1-\alpha)^2} u_i (\mbox{only $i$
learns the secret})\\
&\mbox{\ \ \ } + \frac{(1-\alpha)^2}{\alpha^2 + (1-\alpha)^2}
u_i(\mbox{no one
learns the secret})\\
> &u_i(\mbox{everyone learns the secret}).
\end{array}
\end{equation}
It follows from U1--U3 that
\begin{equation}\label{eq2}
\begin{array}{ll}
&u_i (\mbox{only $i$ learns the secret}) \\
> &u_i(\mbox{everyone learns the secret})\\
> &u_i(\mbox{no one learns the secret}).
\end{array}
\end{equation}
It is immediate from (\ref{eq2}) that there exists some $\alpha^*$
such that, for all $i$ and all $\alpha < \alpha^*$,  (\ref{eq1})
does not hold. Thus, if $\alpha < \alpha^*$, then no player has
any incentive to cheat at step 3.

It is easy to check that each player has no incentive not to send
bits as required in step 1 and 2, assuming that the other players
are following the recommended strategy; this will simply cause the
other players  to stop playing.  Suppose that the bits $c_{(i,+)}$
and $c_{(i,-)}$ that player $i$ actually sends at step 1 are not
the ones that it was supposed to send.  This is easily seen to be
equivalent to player $i$ changing the distribution with which
$c_i$ and $c_{(i,+)}$ are chosen. Although this changes the
probability that messages will be sent in step 3, it is easy to
show that it does not affect the probabilities in (\ref{eq1}).
Thus, player $i$'s expected utility does not change if player $i$
changes the probabilities in step 1, so player $i$ has no incentive
to cheat at step 1 if all other players follow the recommended
strategy.  Finally, it is easy to show that player $i$ will not
cheat at step 2, since this just means that $i^-$ may incorrectly
compute $c_1 \oplus c_2 \oplus c_3$, which may cause the protocol
to terminate with no one learning the secret, and will certainly
not cause $i$ to learn the secret, since at most one of the others
will send its share. This argument shows that the recommended
protocol in $M(\alpha)$ is a Nash equilibrium for $\alpha <
\alpha^*$.

To show that the recommended protocol survives iterated deletion
of weakly-dominated strategies,
consider strategies for player $i$ that the following property:
\begin{itemize}
\item[(*)] if there is nontrivial randomization at node $v$, then player $i$
does not have all $m$ shares and does not send out any shares.
\end{itemize}
Note that the protocol above satisfies (*), so it
suffices to show that strategies satisfying
(*) are not deleted.  This is relatively straightforward, using the
observation that, by
Proposition~\ref{pro:delete}, all deterministic strategies not in
$\E^1 \union \ldots \union \E^4$ survive iterated deletion.  (Note
that, since the game here is infinite, $\B^h$ $\C^h$, $\D^h$, and
$(\D^h)'$ are all empty in this case.)
We leave details to the full paper.

This completes the argument for 3 out of 3 secret sharing.
To do $m$ out of $n$ secret sharing for $m \ge 3$, $n > 3$, we
simply partition the players into three groups, and designate $m$
players such that each group has at least one of the $m$ designated
players.  One of the designated players in each group is taken to be
the leader.
Each of the $m$ designated players sends its share to its group
leader. The three leaders
then essentially use the mechanism sketched above, except that
when they are supposed to send out their share, they send all the
shares of their group to everyone.
Finally, to do $2$ out of $n$ secret sharing for $n \ge 3$, the two
players with shares partition their shares into $n-1$ subshares, and
send the subshares to the other $n-1$ players, along with a zero
knowledge proof that they have constructed the subshares honestly.
Thus, the two players with the
original shares will each have one subshare, while the other players
will have two subshares.  The players then do $n$ out of $n$ secret
sharing, with the subshares being the secrets.
\eprf

There is an important caveat to this result.  The mechanism requires
that each player (or the system designer) knows the other players'
utility functions.   This is necessary in order to choose the
probability $\alpha$ appropriately.  It actually is not critical that
the players know the other players' utility function {\em exactly}.
They just need to know enough about the utility function so as to choose
an $\alpha$
sufficiently small so as
to guarantee that (\ref{eq1}) does not
hold.  In practice, this does not seem unreasonable.

\section{Multiparty Function Computation}\label{sec:multiparty}
As suggested in the introduction, the results for multiparty
computation are similar in spirit to those for secret sharing.
However, some new subtleties arise both in the impossibility
result and the possibility result.

\subsection{The Impossibility Result}\label{sec:multipartyimp}
The impossibility result for multiparty function computation is
essentially a generalization of Theorem~\ref{thm:impsec}. However,
we now no longer want to assume that there is an atomic secret
such that a player's utility depends only on who gets the secret.
Rather, we consider a class of problems where the players have
some initial pieces of ``information'' and the mechanism itself
defines a number of other pieces of information of interest.  A
player's utility again depends on which pieces of information it
and all the other players have at the end of the execution.
Our impossibility result applies
in particular
to the special case of multiparty
computation where the functionality required is the fair exchange
of secrets; as far as we are aware, all available protocols for
this problem have an upper bound on their running time,
and therefore are not appropriate for selfish parties
\cite{Blum83, BN00, Cleve89, Damgard95, EGL, LMR83}.

More formally, assume that there are some pieces of information of
interest, say $I_1, \ldots, I_M$.  In the case of secret sharing,
there is only one piece of information of interest, namely, the
secret.  In the case of multiparty function computation, the
pieces of information of interest are each player's private
information and the value of the function.
A mechanism may define further pieces of information $I_{M+1},
\ldots, I_N$ of interest.   For example, if the secret has $b$
bits, one piece of information may be the first $b/2$ bits of the
secret.

Given a  path $r$ in the game tree, we generalize the notation of
Section~\ref{sec:impossibility1} by letting $\info(r)$ be the
tuple $(\I_1, \ldots, \I_n$), where $\I_i$ is the set of pieces of
information that player $i$ obtains in run $r$.  Let $\info_i(r) =
\I_i$; thus, $\info_i(r)$ is the information that $i$ gets in $r$.

We assume that a player's utility function satisfies the following
analogues of U1--U3:
\begin{enumerate}
\item[V1.] $u_i(r) =  u_i(r')$ if
$\info(r) = \info(r')$.
\item[V2.] If $\info_i(r) \supseteq \info_i(r')$, and $\info_j(r)
\subseteq \info_j(r')$ for $j \ne i$, then $u_i(r) \ge u_i(r')$.
\item[V3.] If
$i \ne j$, $\info_j(r) \subset \info_j(r')$,  $\info_{j'}(r) =
\info_{j'}(r')$ for $j' \ne j$, and $u_j(r) < u_j(r')$, then
$u_i(r) > u_i(r')$.
\item[V4.]  If $\info_i(r_1) = \info_i(r_1')$, $\info_i(r_2) =
\info_i(r_2')$, $\info_j(r_1) = \info_j(r_2)$ for $j \ne i$,
$\info_j(r_1') = \info_j(r_2')$ for $j \ne i$, and $u_i(r_1) <
u_i(r_2)$, then $u_i(r_1') < u_i(r_2')$.
\end{enumerate}
V1 is the obvious generalization of U1; it says that a player's
utility depends only on the information that each of the players
get. V2 says that player $i$ is no worse off if he gets more
information and the other players get less.  V3 says that if
getting more information makes $j$ strictly better off, then $j$
getting that information makes $i$ worse off.  Finally, V4 is an
independence assumption.  It says that whether or not $i$ is
better off with certain information is independent of what the
other players have.  In more detail, since all the players other
than $i$ have the same information in $r_1$ and $r_2$ and
$u_i(r_1) < u_i(r_2)$, then $i$ must be better off with the
information he has in $r_2$ than the information he has in $r_1$.
Since $i$ has the same information in $r_2'$ as in $r_2$, and the
same information in $r_1'$ as in $r_1$, and all other players have
the same information in $r_1'$ and $r_2'$ (although that
information may be different from what they had in $r_1$ and
$r_2$), V4 says that $i$ should also be better off in $r_2'$ than
in $r_1'$.  It is easy to check that if U1--U3 are satisfied for
secret sharing (where the only piece of information of interest is
the secret), then V1--V4 are too. We remark that when a player
``has'' a certain piece of information, there may actually be a
small probability that this information is incorrect.
Nevertheless, as long as the probability is low enough, V1--V4
seem reasonable.

With these assumptions, we can prove an analogue to
Theorem~\ref{thm:impsec} for secret sharing.

\thm\label{thm:impmulti}
If players' utilities satisfy V1--V4, then there is no practical
mechanism $(\Gamma,\vec{\sigma}^*)$ for multiparty function computation
such that $\Gamma$ is finite, and using $\vec{\sigma}^*$, some player
learns the function
value. \ethm

\subsection{A Randomized Practical Mechanism for Multiparty Computation}
In this section we focus on the special case of the framework in the
previous section where the only  pieces of information are $I_0, \ldots,
I_n$, where $I_0$ is the value of the function and $I_i$
is player $i$'s private value, for $i = 1, \ldots, n$.
For simplicity, we restrict attention to
multiparty computation of Boolean functions.%
\footnote{We do this because the results we refer to, in particular
\cite{goldreich03,st:bool}, consider only Boolean functions.  We
believe that we can extend to arbitrary
functions with no difficulty, although we have not worked out the details.}
We say that a run $r$ is {\em admissible} if no player learns any other
player's private information in $r$, and let $\info_{i0}(r)$ be 0 or 1
depending on whether or not $i$ learns the value of the function in $r$.
We assume that the players' utility functions satisfy the following
analogue of U2:
\begin{enumerate}
\item[V5.] If $r$ and $r'$ are admissible, $\info_{i0}(r) = 1$ and
$\info_{i0}(r') = 0$, then $u_i(r) > u_i(r')$.
\end{enumerate}
As we said in the
introduction, it follows from Shoham and Tennenholtz's results
\cite{st:bool} that only certain functions, which they call {\em
non-cooperatively computable} (NCC),
can be computed, even with a trusted party,
because
only for NCC functions do
the players have an incentive to reveal their private inputs
truthfully, if
players' preferences satisfy V1--V5.
Clearly if a function $f$ is not NCC, then we cannot hope to compute its
value with revealing private information in the absence of a trusted
party.  We now show that if it is in NCC, then we can compute its value.

\thm
Assuming the existence of one-way 1-1 functions,
if $f$ is non-cooperatively computable, $n \ge 3$, and
players' utilities satisfy
V1--V5, then there is a (randomized) practical mechanism for
computing $f$ that runs in constant expected time.
\ethm

\prf
The mechanism for multiparty computation combines ideas of the
secret sharing mechanism in Section~\ref{sec:randlength} and the
multiparty function protocol of Goldreich, Micali, and Wigderson
\cite{gmw87}.
Goldreich \citeyear[Section 7.3]{goldreich03} provides a careful proof of the
correctness of the protocol, against what he calls {\em semi-honest\/} or
{\em passive\/} adversaries.  In our context, this means that players
are allowed to lie about their initial values, and are allowed
to abort the protocol for any reason (including getting the value of the
function, in the hopes that other players have not gotten the value),
but all messages they send must be the ones that the protocol says they
should send, given their claimed initial values.
The key
idea is to simulate the computation of a circuit for computing $f$, such
that, at each stage of the protocol, the value of a node in the circuit
is viewed as a secret, and all players have a share of that secret.
Our protocol also uses this circuit simulation, but replaces the last
step of secret sharing (where the players actually learn the value of
the function) by the secret-sharing protocol from the practical
mechanism given in Section~\ref{sec:randlength}.
Goldreich then shows how to force semi-honest behavior, using
zero-knowledge and bit-commitment protocols.  We can employ the same
techniques in our protocol to force semi-honest behavior; this is where
we need the assumption that one-way 1-1 functions exist, just as
Goldreich does.%
\footnote{As we show in the full paper, we can actually simplify Goldreich's
arguments given our assumption of rationality.}
While, as Goldreich points out, in his setting, we cannot force players
to reveal their true input values, if $f$ is NCC, then, by V5, it is to
a player's benefit to reveal the true input value.  Similarly, it is to
the player's benefit not to abort the protocol.

The proof that this strategy is a Nash equilibrium and survives
iterated deletion is
similar to the argument given in the proof of Theorem~\ref{thm:rand},
with the share
issuer implemented by the multiparty computation protocol, and the
zero-knowledge proofs acting in the same way as the signatures on
shares.  We omit details here.
\eprf
\section{Conclusions}\label{sec:conc}
We have shown how to apply ideas of rationality to secret sharing
and multiparty computation.  This allows us to think of players as
being rational, rather than ``good'' or ``bad''.  For many
applications of multiparty function computation, this seems like a
far more reasonable approach.  Many open problems remain,
including the following:
\begin{itemize}
\item We have assumed that we are working in synchronous systems.  We
conjecture that in asynchronous systems there are no practical
mechanisms for secret sharing or multiparty computation satisfying
(some variant of)
U1--U3, but have not yet proved this.
\item We have focused on preferences that satisfy U1--U3.
McGrew, Porter, and Shoham \cite{mps:gen} characterize when
multiparty computation is achievable with a trusted party under
various other assumptions about players' preferences.  We
conjecture that it will continue to be the case that we can do
multiparty computation without a trusted  party whenever we can do
it with a trusted party.
\item We have concentrated on strategies that are Nash equilibria and survive
iterated deletion of weakly-dominated strategies.  What about
other solution concepts?  We note that our impossibility result
(Theorem~\ref{thm:impsec}) does not hold for trembling hand
equilibria, and therefore not for sequential or subgame perfect
equilibria either.
\item Our mechanism requires, not just knowing that the utilities satisfy
U1--U3, but exactly what the utilities are.
Is there
a single mechanism (without a parameter $\alpha$) that
works for any utilities that satisfy U1--U3?
\end{itemize}
\section{Acknowledgments}
We would like to thank
Adam Brandenburger, Cynthia Dwork, Joan Feigenbaum,  Bob McGrew, John
Mitchell, Andrew Postlethwaite, Yoav Shoham, and Moshe Tennenholtz for
helpful discussions about this work.

\bibliographystyle{abbrv}

\begin{thebibliography}{10}

\bibitem{bgw}
M.~Ben-Or, S.~Goldwasser, and A.~Wigderson.
\newblock Completeness theorems for non-cryptographic fault-tolerant
  distributed computation.
\newblock In {\em Proc.~20th ACM Symp.~on Theory of Computing}, pages 1--10,
  1988.

\bibitem{Blum83}
M.~Blum.
\newblock How to exchange (secret) keys.
\newblock {\em ACM Trans. on Computer Systems}, 1(2):175--193, 1983.

\bibitem{BN00}
D.~Boneh and M.~Naor.
\newblock Timed commitments.
\newblock In {\em Proc.~CRYPTO 2000}, Lecture Notes in Computer Science,
  Volume~1880, pages 236--254. Springer-Verlag, 2000.

\bibitem{BK00}
A.~Brandenburger and J.~Keisler.
\newblock Epistemic conditions for iterated admissibility.
\newblock Unpublished manuscript; first version 6/16/00, latest draft 6/9/03.,
  2000.

\bibitem{canetti96studies}
R.~Canetti.
\newblock {\em Studies in Secure Multiparty Computation and Applications}.
\newblock PhD thesis, Technion, 1996.

\bibitem{ccd}
D.~Chaum, C.~Cr\'{e}peau, and I.~Damg{\aa}rd.
\newblock Multi-party unconditionally secure protocols.
\newblock In {\em Proc.~20th ACM Symp.~on Theory of Computing}, pages 11--19,
  1988.

\bibitem{Cleve89}
R.~Cleve.
\newblock Controlled gradual disclosure schemes for random bits and their
  applications.
\newblock In {\em Proc.~CRYPTO '89}, pages 573--588, 1989.

\bibitem{Damgard95}
I.~Damg{\aa}rd.
\newblock Practical and provably secure release of a secret and exchange of
  signatures.
\newblock {\em Journal of Cryptology}, 8(4):201--222, 1995.

\bibitem{EGL}
S.~Even, O.~Goldreich, and A.~Lempel.
\newblock A randomized protocol for signing contracts.
\newblock {\em Communications of the ACM}, 28(6):637--647, 1985.

\bibitem{FHMV}
R.~Fagin, J.~Y. Halpern, Y.~Moses, and M.~Y. Vardi.
\newblock {\em Reasoning about Knowledge}.
\newblock MIT Press, Cambridge, Mass., 1995.

\bibitem{FS02}
J.~Feigenbaum and S.~Shenker.
\newblock Distributed algorithmic mechanism design: Recent results and future
  directions.
\newblock In {\em Proc.~6th International Workshop on Discrete Algorithms and
  Methods for Mobile Computing and Communications}, pages 1--13. ACM Press,
  2002.

\bibitem{goldreich03}
O.~Goldreich.
\newblock {\em Foundations of {C}ryptography, {V}ol. 2}.
\newblock Cambridge University Press, 2004.
\newblock To appear; draft available at
  www.wisdom.weizmann.ac.il/$\sim$oded/foc.html.

\bibitem{gmw87}
O.~Goldreich, S.~Micali, and A.~Wigderson.
\newblock How to play any mental game.
\newblock In {\em Proc.~19th ACM Symp.~on Theory of Computing}, pages 218--229,
  1987.

\bibitem{LMR83}
M.~Luby, S.~Micali, and C.~Rackoff.
\newblock How to simultaneously exchange a secret bit by flipping a
  symmetrically-biased coin.
\newblock In {\em Proc.~24th IEEE Symp.~on Foundations of Computer Science},
  pages 11--21, 1983.

\bibitem{McGrewBGW}
R.~McGrew, R.~Porter, and Y.~Shoham.
\newblock Towards infomational mechanism design: A new perspective on secure
  function evaluation.
\newblock Unpublished manuscript.

\bibitem{mps:gen}
R.~McGrew, R.~Porter, and Y.~Shoham.
\newblock Towards a general theory of non-cooperative computing.
\newblock In {\em Theoretical Aspects of Rationality and Knowledge: Proc.~Ninth
  Conference (TARK 2003)}, pages 59--51, 2003.

\bibitem{OR94}
M.~J. Osborne and A.~Rubinstein.
\newblock {\em A Course in Game Theory}.
\newblock MIT Press, Cambridge, Mass., 1994.

\bibitem{shamir}
A.~Shamir.
\newblock How to share a secret.
\newblock {\em Communications of the ACM}, 22:612--613, 1979.

\bibitem{st:bool}
Y.~Shoham and M.~Tennenholtz.
\newblock Non-cooperative computing: Boolean functions with correctness and
  exclusivity.
\newblock {\em Theoretical Computer Science}, 2004.
\newblock To appear. Also available at
  iew3.technion.ac.il/$\sim$moshet/NCC-TCS.pdf.

\bibitem{yao:sc}
A.~Yao.
\newblock Protocols for secure computation (extended abstract).
\newblock In {\em Proc.~23rd IEEE Symp.~on Foundations of Computer Science},
  pages 160--164, 1982.

\end{thebibliography}

\balancecolumns %
\end{document}